\documentclass{ws-procs9x6}

\begin{document}

\title{WARP: a WIMP double phase Argon detector}

\author{R. Brunetti, E. Calligarich, M. Cambiaghi, C. De Vecchi,\\
R. Dolfini, L. Grandi, A. Menegolli, C. Montanari, M. Prata,\\
A. Rappoldi, G.L. Raselli, M. Roncadelli, \\
M. Rossella, C. Rubbia\footnote{Also at \uppercase{ENEA},  \uppercase{P}residenza,  \uppercase{L}ungotevere  \uppercase{T}haon  \uppercase{D}i  \uppercase{R}evel 76,  \uppercase{I}00196  \uppercase{R}oma,  \uppercase{I}taly} and C. Vignoli}

\address{INFN and Department of Physics at University of Pavia,\\
Via Bassi 6, I-27100 Pavia (PV), Italy}

\author{F. Carbonara, A. Cocco, A. Ereditato, G. Fiorillo,\\
G. Mangano and R. Santorelli }

\address{INFN and Department of Physics at University of Napoli,\\
Via Cintia, I-80126 Napoli (NA), Italy}

\author{F. Cavanna, N. Ferrari and O. Palamara}

\address{INFN Gran Sasso Laboratory,\\
INFN, Laboratori Nazionali del Gran Sasso (LNGS),\\ 
S.S. 17/bis Km 18+910, I-67010 L'Aquila - Italy}

\author{Presented by L. Grandi}
\address{E-mail: luca.grandi@pv.infn.it}

\maketitle

\abstracts{
The \emph{WARP} programme for dark matter search with a double phase argon detector is presented. In such a detector both excitation and ionization produced by an impinging particle are evaluated by the contemporary measurement of primary scintillation and secondary (proportional) light signal, this latter being produced by extracting and accelerating ionization electrons in the gas phase. The proposed technique, verified on a $2.3\ liters$ prototype, could be used to efficiently discriminate nuclear recoils, induced by \emph{WIMP's} interactions, and measure their energy spectrum. An overview of the $2.3\ liters$ results and of the proposed $100\ liters$ detector is shown.
}

\section{Introduction}
\label{sec:intro}
The very small interaction cross section makes \textit{WIMP}-nucleus scattering a very rare event. Such a rate is not easily predicted, since it depends on many variables which are poorly defined. In practice, uncertainties may encompass many orders of magnitude, although the minimal \textit{SUSY} leaves open the optimistic possibility of very significant rates \cite{ref:kim}. Any new experiment must therefore reach an ultimate sensitivity which is several orders of magnitudes higher than the one of the presently ongoing searches \cite{ref:dama}. To achieve such a goal, both sensitive mass and background discrimination should be as large as possible. From this point of view the cryogenic noble liquids technology (Ar and Xe) is one of the most promising techniques since it provides both a highly efficient discrimination and the potentiality to be extended to multi-ton sensitive volumes. Already in 1993 \cite{ref:xenon93} the \textit{ICARUS} collaboration pointed out, for the first time, that the simultaneous measurement of scintillation and ionization produced in noble liquids by an impinging particle permits to efficiently discriminate the nature of the particle itself. In particular this makes it possible to discriminate a nuclear recoil, eventually induced by a \emph{WIMP} in the typical energy range $\approx 10-100\ keV$, from the dominant gammas and electrons background induced by materials radioactivity.

The \emph{WARP} (\emph{W}imp \emph{AR}gon \emph{P}rogramme) collaboration has focused on argon since its technology is already fully operational, well supported at industrial level and it has low cost, providing at the same time a sensitivity similar to xenon for a reasonable energy threshold ($E_{rec}>30 keV$): the less effective coherent effect is compensated by the less steeper nuclear form factor \cite{ref:warp}. 

To verify the proposed technique a $2.3\ liters$ double phase argon detector has been realized and an intensive set of measurements and calibrations with various radioactive sources has been conducted (see Sec.~\ref{sec:test}). At the end of this feasibility study the argon chamber, opportunely refurbished, has been installed at the underground Gran Sasso National laboratory (LNGS) to study the background (see Sec.~\ref{sec:LNGS}). On the basis of the obtained results a 100 liters sensitive volume argon detector, to be installed at LNGS, has been proposed \cite{ref:warp}. The construction of the $100\ liters$ detector has been approved and funded by INFN starting from 2004 (see Sec.~\ref{sec:100liters}).

\section{2.3 liters chamber: feasibility study}
\label{sec:test}
The $2.3\ liters$ drift chamber consists of a lower liquid volume and an upper region with Argon in the gaseous phase, kept in thermal equilibrium by immersing the hole chamber in a liquid argon bath. LAr electro-negative impurity concentration is kept at a level $\leq 0.1\ ppb$ ($O_{2}$ equiv.) with the use of standard $Hydrosorb^{TM}$ and $Oxisorb^{TM}$ filters. The prototype is equipped with a single $8$'' cryogenic photomultiplier coated with TPB to wave-shift VUV scintillation photons \cite{ref:warp} and installed in the gaseous phase. In such an experimental setup the ionization electrons, produced in the liquid by an interacting particle, are drifted towards the liquid-gas boundary, extracted with the help of an electric field and accelerated in a high electric field region to produce proportional scintillation light. Extraction and multiplication fields are generated through the use of grids. A high diffusive reflector layer surrounds the inner volume to increase light collection efficiency. The typical light signal generated by an interaction in the liquid volume and recorded by the PMT is then constituted by a prompt primary peak (primary signal $S1$), produced by de-excitations and recombination processes, followed after a drift time (depending on actual location of the interaction) by a secondary peak (secondary signal $S2$), associated to the ionization electrons drifted in the liquid and accelerated in the gas phase.

As a first test the chamber has been used as a standard scintillation counter with extraction and multiplication fields turned off. In this way only primary signal survives since the eventually drifted electrons reaching the liquid-gas boundary do not produce proportional light. To evaluate the collected light yield and hence the detection efficiency,  the chamber has been exposed to a variety of radioactive sources: in particular a $^{109}Cd$ source, providing several x-rays peaks in the region $20-25\ keV$, has been placed inside the sensitive volume. The measured photo-electron yield, deduced from several acquired scintillation spectra, is $\Gamma(0)=2.9\ phe/keV$ at zero drift field and $\Gamma(1000)=2.35\ phe/keV$ at $1\ kV/cm$ drift field: this decrease is due to the fact that the applied drift field avoids ionization electrons-ions recombination.

With a similar setup but the fields on, the electrons extraction process has been studied as function of the extraction field (maintaining a fixed multiplication field). The extraction probability is obviously function of the electric field applied at the liquid-gas interface through the use of two grids, one placed in liquid (\emph{g1}) and the other in gas (\emph{g2}). The measured extraction efficiency is in excellent agreement with experimental results found in literature\cite{ref:gushchin}. At fields of the order of $2.5\ kV/cm$ almost the totality of the electrons reaching the interface is extracted in less than $0.1\ \mu s$ differently from what occurs in xenon for which higher fields are required.

Intensive tests on multiplication process has been conducted too. Providing a fixed extraction field, the amplitude of the secondary signal as function of the applied field has been investigated. Proportional light production starts at fields of the order of $1\ kV/cm$ (significantly lower than the one needed for xenon). At $5\ kV/cm$ the gas gain is about $32\ photons/electron/cm$ for a saturated gas pressure of $1\ bar$ at $87\ K$).

After having verified the possibility of extracting the ionization electrons and producing proportional light emission with a double phase argon detector, the response of the chamber to different kind of particle has been investigated. Since different particles produce excitation and ionization in a different amount, the ratio $S2/S1$ should be function of the nature of the interacting particle. In particular, we have explored the signatures of an hypothetical \textit{WIMP}, exposing the chamber to a $D-T\ 14\ MeV$ neutron generator and to an Am-Be neutron source: fast neutrons scattering elastically on nuclei behave like ``strong interacting'' \textit{WIMP's}, producing nuclear recoils in the energy range of interest. At the working electric fields a strong correlation between primary and secondary signals is evident: differently from gamma-like signals (electrons mediated) for which $S2/S1\gg1$, the observed nuclear recoil events are characterized by $S2/S1\ll 1$ populating a completly different region of the scatter plot $S2/S1$ versus $S1$. The obtained results are in good agreement with the predictions of the so-called Box Model of Thomas and Imel\cite{ref:imel}: due to local density ionization a smaller amount of electrons, if compared to minimum ionizing particles, is able, under the effect of drift field, to escape from recombination (for details see \emph{WARP} proposal\cite{ref:warp}). The measured photo-electron yield for nuclear recoils, deduced from their primary scintillation spectrum, is $0.66\ phe/keV$ (at $1\ kV/cm$) to be compared to $2.9\ phe/keV$ for gammas at zero drift field.

\section{Run in underground laboratory}
\label{sec:LNGS}
The $2.3\ liters$ chamber, opportunely refurbished, has been installed at the Gran Sasso National Laboratory to work in a low background environment. It has been equipped with seven $2\ inch$ cryogenic photo-multipliers and several technical improvements (for a better field uniformity) have been applied.

As a first step the primary scintillation spectrum has been measured and compared to the one acquired at surface level. As expected the gamma spectrum at LNGS ends at $3\ MeV$. The dominant contributions comes from $^{232}Th$ and $^{238}U$ chains and $^{40}K$.

\begin{figure}[t]
\centerline{\epsfxsize=9cm\epsfbox{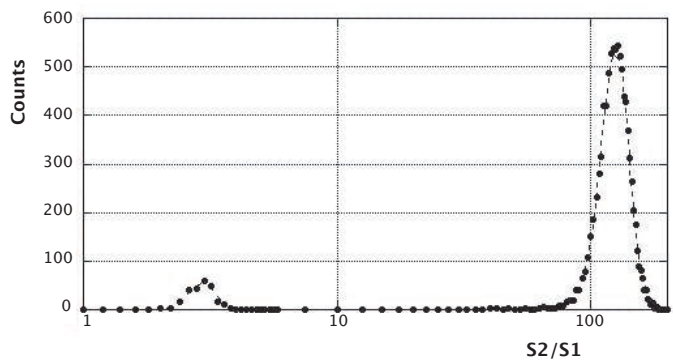}}   
\caption{$S2/S1$ distribution. Two well separated peaks are evident. One due to minimum ionizing particles and the other to $^{222}Rn\rightarrow ^{218}Po +\alpha + 5.489\ MeV$ decays (half-life of $3.825\ days$). }
\label{fig:03}
\end{figure}

With this new internal setup and different fields value a new measurement of the discriminating power has been conduced. Figure \ref{fig:03} shows the $S2/S1$ distribution. Two well separated families are visible: one, associated to minimum ionizing particles (gammas), centered around $125.7$ and another, due to $^{222}$Rn $\alpha$-decay contained in the argon, around $S2/S1=2.9$. The gamma-$\alpha$ suppression factor is about $44/1$. As expected alpha particles, due to their strong recombination, behaves similarly to nuclear recoils providing a depleted ionization signal. Obviously the value of $S2/S1$ is function of the applied electric fields and of the argon gas pressure (equal to the atmospheric pressure and different according to the measurement site).

At the moment the chamber has been surrounded by a $10\ cm$ thick lead shield and it is running in a low background condition to study the effects of shield and internal contamination.

\section{The proposed 100 liters detector}
\label{sec:100liters}
\begin{figure}[t]
\centerline{\epsfxsize=9cm\epsfbox{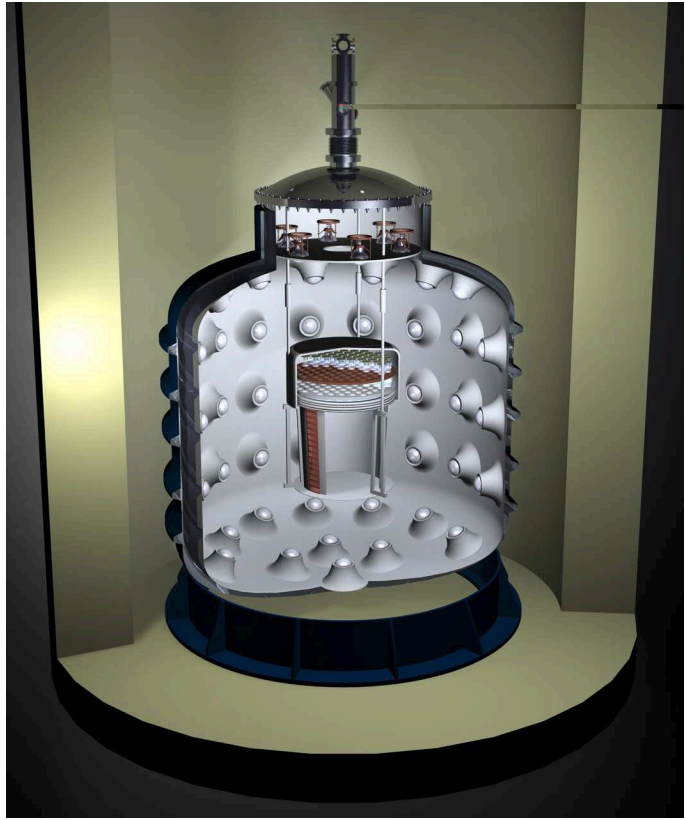}}   
\caption{3D view of the $100\ liters$ prototype. The core of the detector is represented by the double-phase $50\ cm$ drift sensitive volume, completely immersed in a $60\ cm$ thick VETO region readout by $400$ $3\ inch$ photo-multipliers. Through the use of a stainless steel cap and some heaters a gas pocket is maintained in the central volume equipped with $61$ $2\ inch$ PMTs. The external passive shield is also shown.}
\label{fig:02}
\end{figure}
On the basis of the comfortable results obtained with the small prototype, a new argon based detector has been proposed \cite{ref:warp}. Its basic scheme, shown in Figure \ref{fig:02}, foresees a fiducial volume of LAr (about $100\ liters$), tracing the layout of the $2.3\ liters$ chamber, with a uniform electric field drifting ionization electrons towards a liquid to gas interface. A set of grids with an appropriate voltage arrangement provides then the extraction of ionization electrons from the liquid phase and their acceleration in the gas phase  for the production of the secondary light pulse. A set of photomultipliers placed above the grids sense both the primary scintillation signal in the liquid argon and the delayed secondary pulse in the gas phase. \textit{PMTs} granularity allows reconstruction of event position in the horizontal plane with about ${1\ cm}$ resolution. Position along the drift coordinate is given by the drift time (position reconstructed in 3D). The whole detector has been designed trying to minimize the weight and therefore the amount of materials (and radioactive contamination) to be placed around the inner active volume.

The detector is completely submersed in a LAr volume that works as an anti-coincidence (\textit{Active VETO}), which is also readout by a set of phototubes, the two volumes are optically separated. The \textit{VETO} region is used to reject the events due to neutrons or other particles penetrating from outside or travelling out from the central part. Dimensions of the outer LAr volume are chosen in such a way that the probability for a neutron to interact in the inner detector without producing a signal in the \textit{VETO} system is negligible. Only events with no signals in the \textit{VETO}  are potential \textit{WIMP's} candidates. 

An external shield is added to adequately reduce environmental neutron and gamma background. With ${100\ liters}$ sensitive volume, which corresponds to an active mass of about ${150\ kg}$, thanks to the rejection/identification power of the two-phase technique, we could reach a sensitivity about two orders of magnitude better than the present experimental limit from \textit{CDMS} (or to the presently indicated hint from the \textit{DAMA} experiment).

\end{document}